\documentclass[aps,prl,showpacs,twocolumn,superscriptaddress,longbibliography]{revtex4-1} 	

\usepackage{graphicx, amsmath, amssymb}
\usepackage{epstopdf}		

\usepackage{multirow}
\usepackage{dcolumn}
\newcolumntype{d}{D{.}{.}{1}}

\def \lvec{(\kern-.26em(}

\begin{document}

\title{Control of Optical Transitions with Magnetic Fields in Weakly Bound Molecules}
%
%
\author{B. H. McGuyer}
\author{M. McDonald}
\author{G. Z. Iwata}
\affiliation{Department of Physics, Columbia University, 538 West 120th Street, New York, New York 10027-5255, USA}
\author{W. Skomorowski}
\altaffiliation{Present address: Theoretische Physik, Universit\"{a}t Kassel, Heinrich Plett Stra{\ss}e 40, 34132 Kassel, Germany}
\author{R. Moszynski}
\affiliation{Department of Chemistry, Quantum Chemistry Laboratory, University of Warsaw, Pasteura 1, 02-093 Warsaw, Poland}
\author{T. Zelevinsky}
\email{tz@phys.columbia.edu}
\affiliation{Department of Physics, Columbia University, 538 West 120th Street, New York, New York 10027-5255, USA}
\date{\today}%

\begin{abstract}
In weakly bound diatomic molecules, energy levels are closely spaced and thus more susceptible to mixing by magnetic fields than in the constituent atoms.  We use this effect to control the strengths of forbidden optical transitions in $^{88}$Sr$_2$ over 5 orders of magnitude with modest fields by taking advantage of the intercombination-line threshold.  The physics behind this remarkable tunability is accurately explained with both a simple model and quantum chemistry calculations, and suggests new possibilities for molecular clocks.  We show how mixed quantization in an optical lattice can simplify molecular spectroscopy.  Furthermore, our observation of formerly inaccessible $f$-parity excited states offers an avenue for improving theoretical models of divalent-atom dimers.

\end{abstract}

\maketitle

Transitions between quantum states are the basis for spectroscopy and the heart of atomic clocks.
The ability to access a transition experimentally depends on the transition mechanism and the states involved.
For atoms and molecules, the dominant mechanism is the electric-dipole interaction, and electric-dipole transitions are only allowed between angular-momentum eigenstates with opposing parity that satisfy the rigorous selection rules
$\Delta J\equiv J'-J=0, \pm 1$ and $\Delta m\equiv m'-m = 0, \pm1$ (but $\Delta J \neq 0$ if $J=0$),
where $J$ and $m$ are the total and projected angular momentum quantum numbers, and primes refer to the higher-energy states.
Accessible transitions that are forbidden
by these rules or the additional rules that arise, for example, from molecular symmetries,
are of great interest because they are associated with long-lived quantum states and enable precision measurements such as parity-violation experiments \cite{guena:2005,derevianko:2007,dzuba:2012,cahn:2014}.
Forbidden transitions are central to atomic time keeping and have been extensively researched in order to advance the state of the art \cite{DereviankoRMP11_OpticalLatticeClocksReview,vanier:1989}.

In this Letter, we demonstrate how the control of forbidden transitions with applied magnetic fields is greatly enhanced by the dense level structure of molecules as compared to atoms.
We use modest fields of a few tens of gauss to not only enable strongly forbidden transitions, but yield transition strengths comparable to allowed transitions.
In contrast, several million gauss would be needed to achieve the same results using the atoms that form these molecules.
The physics that enables this tuning of transition strengths by 5 orders of magnitude also leads to highly nonlinear Zeeman shifts which we precisely measure.
We explain our observations with an intuitive as well as a rigorous theoretical model, and suggest how they may be used to improve such models and to engineer an optical molecular clock.

\begin{figure}
	\centering
	\includegraphics[]{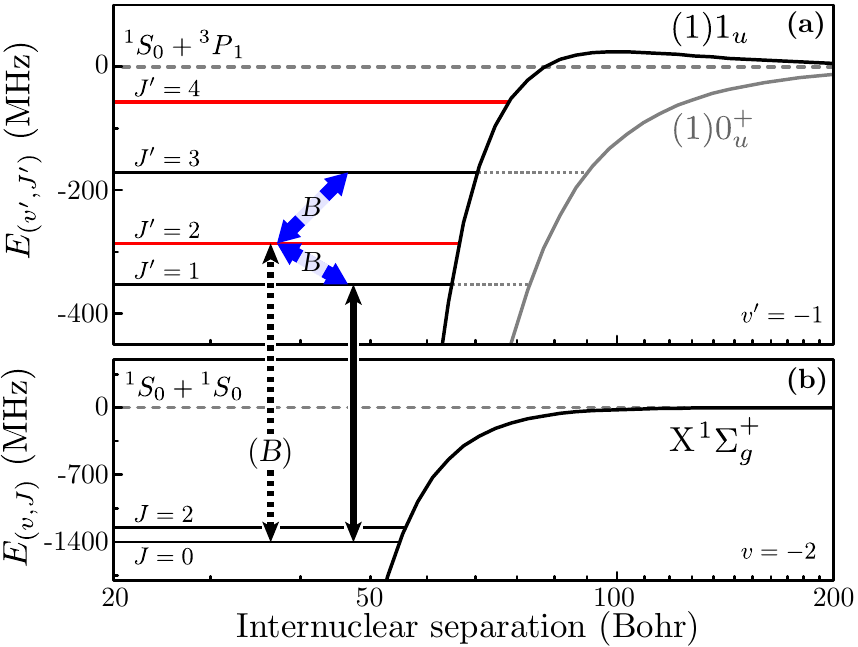}
	\caption{
A magnetically enabled forbidden molecular transition.
(a) Admixing of excited states by an applied static magnetic field $B$ (slanted arrows).
Near the $^1S_0+{^3P_1}$ asymptote, two molecular potentials, $1_u$ and $0_u^+$, are optically accessible from the ground state, X$^1\Sigma_g^+$.
States with odd $J'$ are of both $1_u$ and $0_u^+$ character because of nonadiabatic Coriolis coupling \cite{mcguyer:2013}. This coupling is not essential to this work, and only admixing of $1_u$ states is shown for clarity.
(b) An optical transition from $J=0$ to $J'=2$ is forbidden (dashed vertical arrow), while to $J'=1$ is allowed (solid vertical arrow).
With an applied field, the forbidden transition becomes allowed because of admixing with the $J'=1$ state. 
}
	\label{fig1}
\end{figure}
Figure \ref{fig1} illustrates the process of magnetically enabling a forbidden transition in $^{88}$Sr$_2$ molecules near the atomic $^1S_0 - {^3P}_1$ intercombination line.
While the physics responsible for the effect is not unique to $^{88}$Sr$_2$, this narrow ($\sim10$ kHz) optical transition allows us to (i) spectroscopically address very weakly bound molecules where the energy level density and the magnetic moment are large, and (ii) trap and probe the molecules in an optical lattice without spectral broadening and thus attain sensitivity to tiny transition strengths.
Starting from a ground state with $J=0$, a transition to an excited state with $J'=1$ is allowed (solid arrow in Fig.~\ref{fig1}).  A transition to $J'=2$, in contrast, is forbidden (dashed arrow).
However, applying a static magnetic field couples the excited states (blue arrows), making the energy eigenstates no longer angular-momentum eigenstates.
Thus, the excited state originally described by $J'=2$ acquires a $J'=1$ component that now satisfies the selection rules for a transition from $J=0$.
In this way, applying a magnetic field enables the forbidden transition with $\Delta J = 2$.

\begin{figure}
	\centering
	\includegraphics[width=0.9\columnwidth]{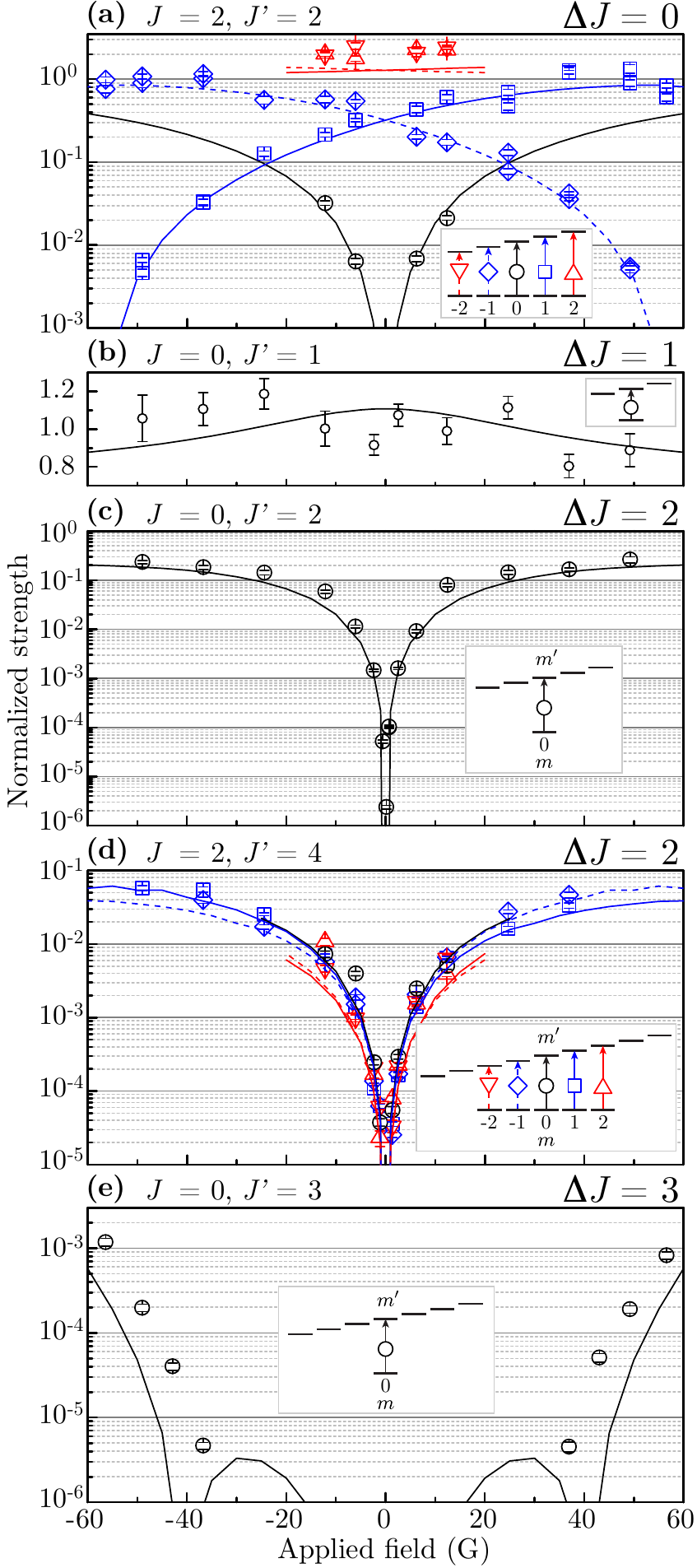}
	\caption{
Magnetic control of molecular transitions in $^{88}$Sr$_2$ near the intercombination line.
Points are experimental values and curves are theoretical calculations.
The $\pi$ transitions are between X$^1\Sigma_g^+(v=-2,J,m)$ ground states and $1_u(v'=-1,J',m')$ excited states for $J'=1,2,4$, or the $0_u^+(v'=-3,J',m')$ excited state for $J'=3$ \cite{vNegative}.
(a) An allowed transition with $\Delta J=0$ has an "accidentally" forbidden $m'=0$ component that becomes allowed with field, and $m'=\pm1$ components that show field-induced interference from admixing.
(b) An allowed transition with $\Delta J = 1$ is mostly field insensitive.  Its average value is used to normalize the data.
(c,d) Forbidden transitions with $\Delta J = 2$ and strengths that vary over 5 orders of magnitude to
become comparable to allowed transitions.
(e) A highly forbidden transition with $\Delta J = 3$ enabled by second-order admixing.
}
	\label{fig2}
\end{figure}
We measured this variation of transition strengths with an applied magnetic field $B$ for ultracold $^{88}$Sr$_2$ in an optical lattice.
The experimental apparatus follows Refs. \cite{mcguyer:2013,mcguyer:1g}.
The results are arranged by increasing $|\Delta J|$ in Fig.~\ref{fig2}.
As shown, moderate magnetic fields are able to strongly control the strength of transitions between ground- and excited-state molecules near the intercombination line.
We are able to drive forbidden transitions with $|\Delta J|$ up to 3 and to control transition strengths over 5 orders of magnitude to nearly reach the allowed transition strengths.

Our data are supported by the theoretical calculations shown in Fig.~\ref{fig2} (curves).
Qualitatively, we explain these observations as follows.
Consider a transition between a ground state $|\gamma\rangle$ and an excited state $|\mu\rangle$.
The strength of this transition is proportional to the square $|\Omega_{\gamma \mu}|^2$ of the Rabi frequency $\Omega_{\gamma \mu} = \langle \gamma | H_{e} | \mu \rangle/\hbar$, where $H_{e}$ is the electric-dipole interaction Hamiltonian and $\hbar$ is the reduced Planck constant.
Applying a static magnetic field perturbs the states and thus the strength of the transition.
To first order in the field strength $B$,
the excited state becomes
\begin{align}	\label{admixing}
|\mu(B) \rangle \approx |\mu(0)\rangle + \sum_{\nu\neq\mu} (B / B_{\mu \nu}) \, |\nu(0)\rangle,
\end{align}
where the characteristic magnetic fields
$B_{\mu\nu} = (E_\mu - E_\nu)/\langle \mu(0) | H_Z/B | \nu(0) \rangle$
give the admixing per unit $B$ for the pairs of states with energies $E_\mu$ and $E_\nu$,
and the sum is over all states that couple to $|\mu\rangle$ via the Zeeman interaction $H_Z=\mu_B(g_L{\bf L}+g_S{\bf S})\cdot{\bf B}$ \cite{mcguyer:2013}.
The field ${\bf B} = B \hat{z}$ defines our quantization axis, and $H_Z$ couples states with $\Delta m = 0$ and $\Delta J = 0, \pm 1$ (but $\Delta J \neq 0$ if $J=0$).  We assume $|\gamma(B)\rangle \approx |\gamma(0)\rangle$ because spinless $^{88}$Sr$_2$ molecules in the electronic ground state interact very weakly with the magnetic field.

As a result, the strength of the transition changes with the applied field as
\begin{align}	\label{ToyModel}
|\Omega_{\gamma \mu}(B)|^2 \approx |\Omega_{\gamma \mu}(0)|^2
	+ B^2 \left| \sum_{\nu\neq\mu} \frac{\Omega_{\gamma \nu}(0)}{B_{\mu \nu}} \right|^2 \nonumber \\
	+ B \, \sum_{\nu\neq\mu} \left( \frac{{\Omega_{\gamma \mu}(0)\Omega_{\gamma \nu}^*(0)}}{B_{\mu\nu}^*} + \frac{{\Omega_{\gamma \mu}^*(0)\Omega_{\gamma \nu}(0)}}{B_{\mu\nu}} \right).
\end{align}
For a forbidden transition, the first and last terms are zero, so the strength will be quadratic in $B$ if $|\mu\rangle$ admixes with a state $|\nu\rangle$ for which the transition would be allowed.
This is what we observe at low fields in Figs.~\ref{fig2}(c,d) and, additionally, in Fig.~\ref{fig2}(a) for the "accidentally" forbidden $m=m'=0$ component of an allowed transition \cite{varshalovich}.
For allowed transitions, all the terms in Eq.~(\ref{ToyModel}) may contribute.
The first term is field insensitive and dominates in Fig.~\ref{fig2}(b).
The third term is linear with $B$ and represents the destructive or constructive interference that we observe with $m=m'=\pm1$ components in Fig.~\ref{fig2}(a).
Finally, the behavior of the highly forbidden transition in Fig.~\ref{fig2}(e) is roughly quartic with $B$, and comes from higher-order admixing beyond this approximate model.

\begin{figure}
	\centering
	\includegraphics[width=\columnwidth]{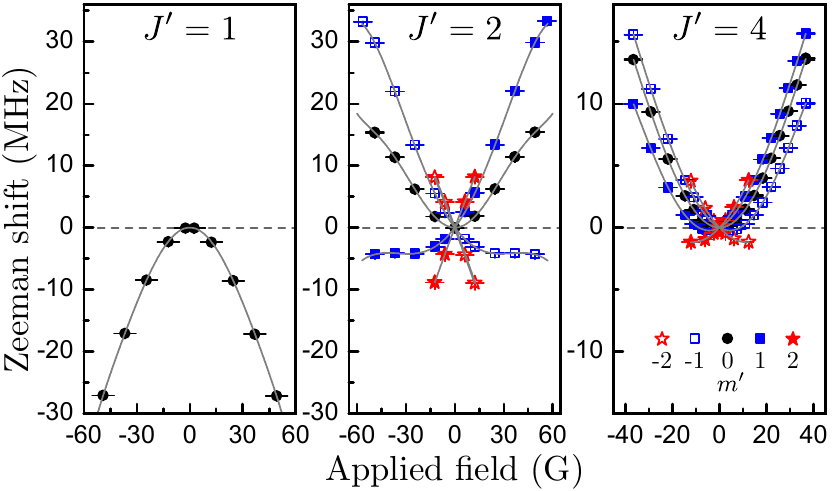}
	\caption{
Highly nonlinear Zeeman shifts for the three $1_u(v'=-1,J')$ states listed in Table 1.  The $\pi$ transitions used in the measurements are indicated in Figs. \ref{fig2}(b), \ref{fig2}(a), \ref{fig2}(d), respectively.
The lines are polynomial fits using Eq. (\ref{ZeemanFit}) with appropriate symmetry constraints.
}
	\label{fig3}
\end{figure}
\begin{table*}
\caption{\label{tab:2}
Experimental (Expt.) and theoretical (Th.) Zeeman shifts for the $1_u(v'=-1)$ states shown in Figs.~\ref{fig2} and \ref{fig3}.
Binding energies $E_b$ are reported to the nearest MHz.
Only the parameters $q_n = q_n(v',J',|m'|)$ (G$^{1-n}$) required for a good fit with Eq. (\ref{ZeemanFit}) are reported.
Coefficients for the $0_u^+(v'=-3,J'=3)$ state in Fig. \ref{fig2}(e) with $E_b = 132$ MHz are available in Ref. \cite{mcguyer:2013}.
}
\begin{ruledtabular}
\begin{tabular}{l c c l l c l l r r r r r r r r}
{$J'$}	& \multicolumn{2}{c}{$|E_b|$} 	
				& \multicolumn{2}{c}{$g$} 	
				& {$|m'|$} 	& \multicolumn{2}{c}{$q_2 \times 10^2$}
							& \multicolumn{2}{c}{$q_3 \times 10^5$}
								& \multicolumn{2}{c}{$q_4 \times 10^6$}	
									& \multicolumn{2}{c}{$q_5 \times 10^{9}$}
										& \multicolumn{2}{c}{$q_6 \times 10^{10}$}	\\
	& (Expt.)	& (Th.) 	
				& (Expt.)	& (Th.)	
				& 	 	& (Expt.)	& (Th.)
						& (Expt.)	& (Th.)
							& (Expt.)	& (Th.) 	
								& (Expt.)	& (Th.) 	
									& (Expt.)	& (Th.) 	\\
\hline
1 	& 353 	& 353	& 0.625(9)	
					& 0.613	
				& 0		& $-$1.122(4) 
						& $-$1.11
							& 0		
							& 0
								& 2.0(1)		
								& 1.85	
									& 0		
									& 0		
										& $-$2.74(2)	
										& $-$2.48		\\
		& & & & & 1 		& $-$0.8(1)  	
						& $-$0.67  	\\
2	& 287 	& 288	& 0.2479(2)	
					& 0.250	
				& 0 		& 0.872(6) 
						& 0.827	
							& 0 	
							& 0 		
								& $-$2.38(6) 	
								&  $-$2.35	
									& 0 	
									& 0			
										& 2.7(1) 	 
										& 2.66	\\ 
		& & & & & 1		& 0.599(1) 	
						& 0.578 
							& 1.1(1)  	
							& 1.7
								& $-$0.97(1) 	
								& $-$1.00
									& $-$5.6(6) 		
									&  $-$6.06	
										& 			 
										& 			\\
		& & & & & 2		& $-$0.18(1)		
						& $-$0.20		\\
4 	& 56 		& 61		& 0.0734(2)
					& 0.075	
				& 0		& 0.882(1)		
						& 0.884	
							& 0			
							& 0	
								& $-$1.16(1)		
								& $-$1.13	\\
		& & & & & 1 		& 0.831(1)		
						& 0.827
							& $-$2.4(1) 	
							& $-$2.48
								& $-$1.09(1)	
								& $-$1.02
									& 7.4(6) 		
									& 7.53 \\
		& & & & & 2		& 0.62(1)		
						& 0.64		\\
\end{tabular}	
\end{ruledtabular}
\end{table*}
Besides affecting transition strengths, the applied field produces highly nonlinear Zeeman shifts of the excited states, as shown in Fig.~\ref{fig3}.
We observe shifts up to sixth order in $B$, well beyond the quadratic shifts reported previously for $^{88}$Sr$_2$ or similar dimers \cite{mcguyer:2013,borkowski:2014,kahmann:2014}, and find good agreement with calculations as shown in Table I.
We parametrize these shifts as the sum of linear and nonlinear terms \cite{mcguyer:2013}
\begin{align}	\label{ZeemanFit}
\Delta E_b = g(v',J') \mu_B m' B + \sum_{n>1} q_n(v',J',m') \mu_B B^n,
\end{align}
where $\mu_B$ is the Bohr magneton.
Here, the binding energies $E_b$ are negative, so positive shifts make molecules less bound.
The sum extends over the fewest terms needed to summarize the data, following the symmetry $\Delta E_b(-m',-B) = \Delta E_b(m',B)$.
We used pure $\sigma$ transitions to measure the signs of $g(v',J')$.

The dense level structure of $^{88}$Sr$_2$ molecules allows the observation of these effects near the intercombination line with significantly lower fields than would be needed for $^{88}$Sr atoms.
In atoms, admixing occurs between $^3P_{J'}$ fine structure levels with spacings $|E_\mu - E_\nu|/h$ of several THz \cite{taichenachev:2006,barber:2006}.
For molecules, admixing occurs between rovibrational levels
near the $^1S_0+{^3P_1}$ threshold, with similar magnetic moments but typical spacings of several tens of MHz.
As a result, the characteristic mixing fields $|B_{\mu\nu}| \sim |E_\mu-E_\nu|/\mu_B$ are roughly
several million gauss
for atoms versus several tens of gauss for molecules.
While these fields may be greatly reduced by choosing an atom with hyperfine structure \cite{metcalf:1979} instead of $^{88}$Sr, the enhancement with molecules versus atoms will still be present.
The enhancement would decrease, however, for more deeply bound molecules as the rovibrational spacings increase.
Similar enhancement is expected with Stark-induced transitions using electric fields \cite{budker:book} as are often used in parity-violation experiments \cite{guena:2005,derevianko:2007,dzuba:2012,cahn:2014}.

We obtained the data in Figs.~\ref{fig2} and \ref{fig3} using procedures similar to those in Refs.~\cite{mcguyer:2013,mcguyer:1g}.
For transition strengths, the measured quantity is $Q\equiv A / (\tau P) = |\Omega_{\gamma\mu}(B)|^2/(4 P)$ \cite{Supplemental}, where $A$ is the Lorentzian area of the natural logarithm of an absorption dip, $\tau$ is the probe exposure time, and $P$ is the probe beam power.
The quantity $Q = Q(m,m')$ was measured separately for each transition component between initial $m$ and final $m'$ quantum numbers, and at different applied fields $B$, by observing the loss of ground-state
molecules by absorption.
The final values of $Q$ were normalized to the average strength of the allowed transition in Fig.~\ref{fig2}(b).

\begin{figure}
	\centering
	\includegraphics[width=\columnwidth]{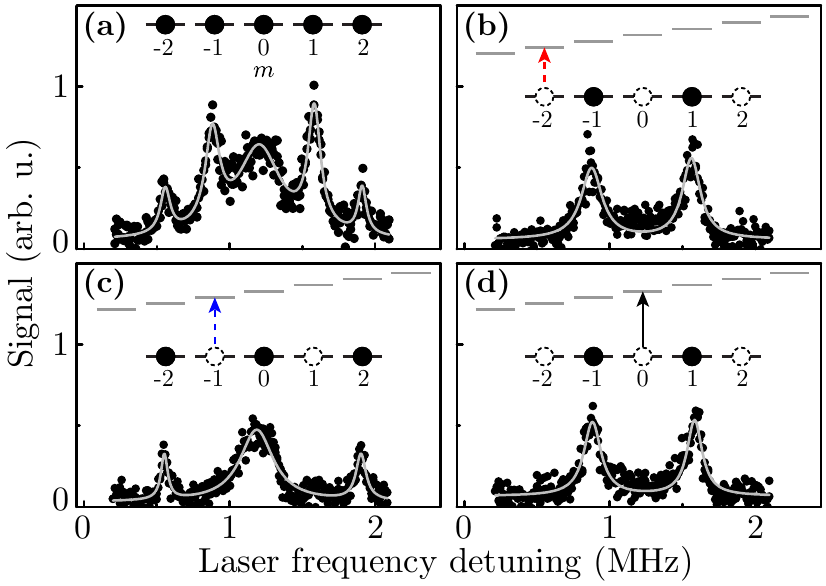}
	\caption{
Demonstration of mixed quantization of $J=2$ ground states by an optical lattice linearly polarized orthogonally to the applied magnetic field.
(a) Spectrum measuring the populations of ground-state sublevels $m$.
(b) A $\pi$ transition to the $m'=-2$ sublevel of a Zeeman-resolved excited state depletes not only the population with $m=-2$, but also with $m=0$ and $m=2$.
The additional population loss with $|\Delta m| = 2,4$ is highly forbidden by selection rules, but occurs because the optical lattice mixes the sublevels.
(c) Likewise, a $\pi$ transition to $m'=-1$ removes the populations with $m=\pm1$.
(d) A $\pi$ transition to $m'=0$ has the same effect as that to $m'=-2$ shown in (b).
}
	\label{fig4}
\end{figure}
To overcome the challenges of quantum-state resolved molecular spectroscopy, we utilized a mixed quantization of the $J=2$ ground-state molecules from competing Zeeman shifts and tensor light shifts \cite{budker:OPA,Supplemental}.
Figure \ref{fig4} demonstrates this effect, showing how the depletion of a sublevel $m$ leads to the depletion of other selected sublevels, simulating forbidden transitions with $|\Delta m|=2,4$.
This effect arises because the optical lattice has an electric field ${\bf E}(t) = E(t) \hat{y}$ linearly polarized orthogonally to ${\bf B} = B \hat{z}$ and to the lattice axis $\hat{x}$ (i.e., the "magic" trapping conditions for a $^1S_0$ -- $^3P_1$ transition in $^{88}$Sr atoms \cite{ido:2003}).
The lattice light shift is therefore not diagonal along ${\bf B}$, but includes off-diagonal couplings (or "Raman coherences" \cite{ido:2003}) between sublevels with $|\Delta m| = 2$.
While the Zeeman shifts (\ref{ZeemanFit}) of the excited states are large enough to suppress these couplings for typical values of $B$, the couplings are large enough to suppress the small Zeeman shifts of the ground states that are only on the order of a nuclear magneton \cite{brown:book}.
As a result, the $J=2$ ground eigenstates are superpositions of sublevels with even or odd $m$, as observed in Fig.~4.
To correct for these effects, the data in Fig.~2 for transitions starting from $J=2$ were multiplied by a correction factor $R(m)$ after normalization, where $R(0) = 4/3$, $R(\pm1) = 2$, and $R(\pm 2) = 8$, as derived in Ref. \cite{Supplemental}.
The mixed quantization enabled our measurement protocol because it provided molecules with all $m$ for $J=2$, in particular, $m=\pm2$ that would be otherwise difficult to create simultaneously.
Furthermore, to measure transition strengths, we only needed to count the final population in two ground-state sublevels ($m=0$ and $m = 1$ or $-1$) to gather the data in Fig.~2.
This was critical at large fields because of the difficulty in individually detecting $m=\pm2$ using the transitions available to convert molecules to atoms.

The theoretical model used the most recent electronic potentials for the
$1_u$ and $0_u^+$
excited states of $^{88}$Sr$_2$ \cite{borkowski:2014}, which are based on the original \textit{ab initio} calculations \cite{MoszynskiSkomorowskiJCP12_Sr2Dynamics}, and the empirical potential for the
ground state \cite{TiemannSteinEPJD10_Sr2XPotential}.
To reproduce the experimental observations of Zeeman shifts, nine excited-state coupled channels including $J'$ = 1 to 6 were required.
Because of the sensitivity to the coupling between the channels, precise measurements of high-order Zeeman shifts as in Table I are useful to test the accuracy of theoretical models \cite{mcguyer:2013,kahmann:2014,borkowski:2014}.
The calculated coefficients $q_2$ and $q_3$ are due to admixing from states with $|\Delta J'| \leq 1$, $q_4$ and $q_5$ with $|\Delta J'| \leq 2$, and $q_6$ with $|\Delta J'| \leq 3$.
The signs of $q_2$ for even and odd $J'$ are typically opposite because of repulsive second-order perturbative couplings between pairs of states, which we have observed for more states than reported here (Fig. \ref{fig3} and Ref. \cite{mcguyer:2013}).

Our direct observation of $1_u$ levels with even values of $J'$ suggests a way to further adjust theoretical models for homonuclear dimers of divalent atoms.
These "$f$-parity" levels are inaccessible by $s$-wave photoassociation, and have not been observed previously in experiments with Sr, Yb, or Ca atoms at ultracold temperatures.
In contrast, they are accessible in experiments with ultracold molecules.
Rovibrational levels with even values of $J'$ exist only for the $1_u$ potential, so Coriolis coupling, which mixes $1_u$ and $0_u^+$ states for odd $J'$, is absent for levels with even $J'$.
Indeed, Table I shows that the $1_u$ levels with even $J'$ have nearly ideal Hund's case (c) $g$ factors \cite{mcguyer:2013},
$g \approx 3/[2J'(J'+1)]$, while those with odd $J'$ do not.
Therefore, precise knowledge of the $1_u$ levels with even $J'$ will allow these potentials to be adjusted independently.

The field enabling of strongly forbidden optical transitions demonstrated here could be used to access ultranarrow molecular transitions.
Particularly, magnetic tuning of transition strengths to long-lived weakly bound subradiant excited states \cite{mcguyer:1g} could enable subhertz optical transitions to $0_g^+$, possibly between a pair of spinless $J=0$ states \cite{taichenachev:2006,barber:2006}.  [The $0_g^+$ and $1_g$ potentials are omitted from Fig. \ref{fig1}(a) due to their extremely weak coupling to the ground state \cite{mcguyer:1g}.]  Clocks based on molecules can complement atomic clocks, for example via different sensitivities to fundamental constant variations \cite{ZelevinskyPRL08,SchwerdtfegerBeloyPRA11_AlphaVarInSr2}.

In conclusion, we have demonstrated the remarkable control of forbidden optical transitions in weakly bound molecules by modest applied magnetic fields.
Our experiments with ultracold $^{88}$Sr$_2$ molecules in an optical lattice are sensitive to exceedingly weak transitions owing to narrow intercombination lines, and demonstrate how mixed quantization can aid molecular spectroscopy as well as suggest new approaches to ultraprecise molecular clocks.  The measurements of transition strengths and highly nonlinear Zeeman shifts provide a stringent test of state-of-the-art quantum chemistry calculations.
The observation of $f$-parity excited-state molecules, in particular, opens new avenues for improvement of future theoretical models for divalent-atom dimers.

We acknowledge NIST Grant No. 60NANB13D163, ARO Grant No. W911NF-09-1-0504, ONR Grant No. N00014-14-1-0802, and NSF Grants No. DGE-2069240 and No. PHY-1349725 for partial support of this work.  R. M. acknowledges the Foundation for Polish Science for the support through the MISTRZ program.

\appendix
\section{Supplemental Material}

\subsection{Mixed quantization in an optical lattice}
To model tensor light shifts of the 1D optical lattice we may use an effective potential of the form
\begin{align}
\label{V}
V 	&= - (1/2) \, \left \langle T^2(\boldsymbol{\alpha}) \cdot T^2({\bf E},{\bf E}) \right \rangle,
\end{align}
where the brackets denote a time average and the spherical tensor notation follows Ref.~\cite{brown:book}.
Using the Wigner-Eckart theorem, this potential evaluates to the matrix
\begin{align}	\label{Vz}
V_{\{|2,m\rangle\}_Z} = \frac{\alpha_2 \, |E|^2}{16}
\begin{pmatrix}
2 & 0 & \sqrt{6} & 0 & 0 \\
0 & -1 & 0 & 3 & 0 \\
\sqrt{6} & 0 & -2 & 0 & \sqrt{6} \\
0 & 3 & 0 & -1 & 0 \\
0 & 0 & \sqrt{6} & 0 & 2
\end{pmatrix}
\end{align}
for a $\{ |J = 2, m = 2\rangle, \ldots, |J = 2, m = -2\rangle \}$ ground-state basis quantized along the applied field ${\bf B} = B\hat{z}$, a lattice electric field ${\bf E} = E \cos(\omega t) \hat{y}$ propagating along $\hat{x}$, and a tensor polarizability
$\alpha_2(\omega) =
	\left(2/\sqrt{105}\right) \left\langle 2 \left|\left| T^2(\boldsymbol{\alpha}) \right|\right| 2 \right\rangle $
that depends on the reduced matrix element of the polarizability operator $\boldsymbol{\alpha}=\boldsymbol{\alpha}(\omega)$ for the particular $J=2$ ground state.
Because ${\bf E} \perp {\bf B}$, the matrix (\ref{Vz}) contains off-diagonal elements from virtual two-photon $\sigma$ transitions that couple pairs of sublevels
$|2,m_1\rangle$ and $|2,m_2\rangle$ with $|m_1 - m_2| = 2$.

The potential (\ref{V}) does not uniquely specify the ground eigenstates because only one eigenenergy is unique.
This is because tensor light shifts are independent of the sign of $m$ for quantization parallel to ${\bf E}$.
Therefore, another interaction is required to break degeneracy and obtain unique eigenstates.
For ground-state $^{88}$Sr$_2$, a likely candidate is the rotational Zeeman interaction, $- g_r \mu_N {\bf J} \cdot {\bf B}$ \cite{brown:book}.
We do not resolve this weak interaction, which is on the order of a nuclear magneton, $\mu_N$, so we will proceed by taking the limit $g_r \rightarrow 0$.
In the basis of matrix (\ref{Vz}), the five unique eigenstates $| \gamma \rangle = \sum_m | J, m \rangle \langle J, m | \gamma \rangle$ obtained from this limit are given by
the following sets of $\langle J, m | \gamma \rangle$ coefficients $\{ \langle 2, 2 | \gamma \rangle, \ldots, \langle 2, -2 | \gamma \rangle \}$:
\begin{align}
\label{MuStart}
\{ 0,1,0,1,0\}/\sqrt{2}, \\
\{ 0,1,0,-1,0\}/\sqrt{2}, \\
\{ \sqrt{3},0,\sqrt{2},0,\sqrt{3}\}/\sqrt{8}, \\
\{ 1,0,-\sqrt{6},0,1\}/\sqrt{8},  \\
\text{and}~\{ 1,0,0,0,-1\}/\sqrt{2}.
\label{MuEnd}
\end{align}
As observed experimentally, these eigenstates consist of mixtures of sublevels with even or odd $m$.

To highlight the mixed nature of this quantization, note that while off-diagonal elements of the tensor light shift (\ref{Vz}) are responsible for mixing if the quantization axis is chosen to be along $\hat{z}$, the mixing would instead come from off-diagonal rotational Zeeman shifts if the quantization axis were chosen to be along $\hat{y}$,
where the potential (\ref{V}) evaluates to $V_{\{|2,m\rangle\}_Y} = - \alpha_2 |E|^2 [3 m^2 - J(J+1)]/[4 J (2J-1)]$.
In this case, the eigenstates would be proportional to the symmetric and antisymmetric combinations $\left( |2,m\rangle \pm |2,-m\rangle \right)$.  
Both cases are equivalent, since the choice of quantization axis is artificial.

\subsection{Measurement of transition strengths}

We measure the strength of a particular Sr$_2$ transition using a procedure similar to that in Ref.~\cite{mcguyer:1g}.
Initially we treat the case without mixed quantization, and return to evaluate its effects in the following section.
First, we apply a laser pulse of duration $\tau$ and power $P$ along the lattice axis to drive the transition between a ground state $|\gamma \rangle$ with quantum numbers $(v,J,m)$ and an excited state $|\mu\rangle$ with $(v',J',m')$.
Afterwards, we measure a signal $S_{J,m}$ proportional to the number $n_{J,m}$ of remaining ground-state molecules with ($v$, $J$, $m$) using a second laser pulse to drive a transition to an excited state that decays to Sr atoms \cite{mcguyer:2013}.
Both laser pulses are linearly polarized along ${\bf B}$ and drive $\Delta m = m' - m =0$ transitions for the spectroscopic quantization axis defined by ${\bf B}$.

We assume the first pulse drives an open transition with sufficiently low power $P$ that the absorption process is linear and any emission back to the initial state is negligible.
During the first pulse, the number of $(v,J,m)$ molecules evolves as
\begin{align}
\label{nEvolution}
\frac{d}{dt} n_{J,m}(t) = -\Gamma_{m, m'}(\delta) \, n_{J,m}(t),
\end{align}
where $\Gamma_{m, m'}(\delta)$ is a stimulated absorption rate per molecule.
The quantity we wish to determine is then
\begin{align}
\label{Qideal}
Q = Q(m,m') = \frac{1}{P} \int \Gamma_{m, m'} (\delta) \, d\delta = \frac{|\Omega_{\gamma\mu}(B)|^2}{4 P},
\end{align}
which is a measure of the strength of the $(m,m')$ component of the transition under study, expressed in terms of the Rabi frequency for this component introduced before Eq.~(1).
This result follows in the rate-equation regime where the evolution (\ref{nEvolution}) applies, because we may use Eq.~(31) of Ch.~4 and Eqs.~(23,32) of Ch.~5 in Ref.~\cite{siegman:book} to write the integrand as
$\Gamma_{m,m'}(\delta) = |\Omega_{\gamma\mu}(B)|^2/ \{\Delta \omega_{12}[1+ (4 \pi \delta)^2 / \Delta\omega_{12}^2]\},$
where $\Delta \omega_{12}$ is the transition linewidth.
We have confirmed the result (\ref{Qideal}) for optical magnetic-dipole transitions to subradiant $1_\text{g}$ excited states \cite{mcguyer:1g}.
Note that the factor of $1/4$ depends on the convention used to define $P$, on $\Delta m$, and on the laser polarization.
For $J=0$, additional relations to convert $Q$ to an Einstein $B$ coefficient and an absorption oscillator strength are available in Ref.~\cite{mcguyer:1g}.

Experimentally, we record the signal $S_{J,m}(\delta) \propto n_{J,m}(\tau) = n_{J,m}(0) \exp[ - \Gamma_{J,m}(\delta) \tau]$ as a function of the laser-frequency detuning $\delta$ from resonance, which is roughly a Lorentzian absorption dip with a constant background.
The quantity $Q$ of Eq. (\ref{Qideal}) is then given by
\begin{align}
\label{Qexpt}
Q = - \frac{1}{\tau P} \int \ln \left[ \frac{S_{J,m}(\delta)}{S_{J,m}(\infty)} \right] d\delta = \frac{A}{\tau P},
\end{align}
where $A$ is the area of a Lorentzian fitted to a plot of $\ln [S_{J,m}(\delta)]$ versus $\delta$, and the shorthand $S_{J,m}(\infty) \propto n_{J,m}(t=0)$ is the signal far from resonance.
No adjustments are made to account for the angular momentum quantum numbers of the initial or final states of the transition.
Care is taken to experimentally verify that $Q$ is independent of $P$ and $\tau$, as expected from Eq. (\ref{Qideal}) because $|\Omega_{\gamma \mu}(B)|^2 \propto P$.
In particular, we use low powers $P$ to avoid power broadening and use pulse times $\tau$ that are short enough to avoid additional loss processes (e.g. collisions) but long enough to avoid transform-limited broadening.

\subsection{Correction for mixed quantization}

For $J=2$ ground states, we follow the same procedure to measure transition strengths as with $J=0$.
However, because of mixed quantization $m$ is no longer a good quantum number.
Instead, the ground eigenstates $|\gamma\rangle$ are superpositions of $|J=2, m\rangle$ sublevels quantized along ${\bf B}$, as derived above.
As a result, the quantity we measure is no longer the $Q$ of Eqs.~(\ref{Qideal}--\ref{Qexpt}), but instead a modified quantity $Q_\text{mix}$.
To account for mixed quantization and obtain the values of $Q$ we report, we multiply these values by a correction factor $R(m_1,m_2)$, as in
\begin{align}
\label{DefineR}
Q(m_1,m') = R(m_1, m_2) \, Q_\text{mix}(m_1,m',m_2),
\end{align}
which we derive as follows.
Here, the two quantum numbers $m_1$ and $m_2$ are determined by the first and second laser pulses, respectively, and defined for quantization along ${\bf B}$.
Because of mixed quantization, these two quantum numbers need not be equal.

Let us denote the number of molecules in the state $|\gamma\rangle$ by $M_\gamma(t)$.
For typical applied fields, the Zeeman interaction of the $J'$ excited state ensures that $m'$ is a good quantum number and that each excited-state sublevel is spectroscopically resolvable.
Therefore, we can tune the frequency and polarization of the first laser to only allow absorption by the ground-state sublevel $|J=2, m_1\rangle$ quantized along ${\bf B}$ (specifically, $m_1 = m'$).  During the first laser pulse, the evolution (\ref{nEvolution}) is then replaced by
\begin{align}
\frac{d}{dt} M_{\gamma}(t) = -\Gamma_{m_1,m'}(\delta) f_\gamma (m_1) M_{\gamma}(t),
\end{align}
where the mixing coefficients
\begin{align}
\label{fmu}
f_\gamma(m) = | \langle J=2, m| \gamma \rangle |^2
\end{align}
may be computed using expressions ({\ref{MuStart}--\ref{MuEnd}).
The number of molecules remaining afterwards is then
\begin{align}
M_\gamma(\tau) = M_\gamma(0) \exp \left[- f_\gamma(m_1) \Gamma_{m_1,m'}(\delta) \tau \right].
\end{align}
Likewise, we can tune the second laser to only detect molecules in the sublevel $|J=2,m_2\rangle$ quantized along ${\bf B}$.
The signal we measure is then
\begin{align}
\label{Smixed}
S_{2,m_2}(\delta) \propto \sum_{\gamma} f_\gamma(m_2) M_\gamma(\tau).
\end{align}
Far off resonance we will denote this by $S_{2,m_2}(\infty) \propto \sum_{\gamma} f_\gamma(m_2) M_\gamma(0)$.

The initial numbers $M_\gamma(0)$ depend on the molecule creation procedure, which consists of photoassociation to an excited state $(J'=1,m'=0)$ followed by spontaneous decay to ground states with $J=0$ and 2 (symmetry forbids ground states with odd $J$) \cite{mcguyer:2013,reinaudi:2012}.
For our experiment, they are
\begin{align}
M_\gamma(0) = \sum_{m} f_\gamma(m) \, n_{2,m}(0),
\end{align}
where the initial numbers $n_{2,m}(0)$ for the unmixed sublevels $|J=2,m\rangle$ are given by the branching ratios
\begin{align}
\label{n2m}
n_{2,m}(0) = n_2(0) \sum_{q=-1}^1 \left( C^{1,0}_{2,m;1,q} \right)^2,
\end{align}
expressed here with Clebsch-Gordon coefficients \cite{varshalovich}.
Note that there would be no initial molecules with $m = \pm 2$ without mixed quantization because of the $\Delta m = 0, \pm1$ selection rule, as in $n_{2,\pm2}(0) = 0$.
However, because of mixed quantization there will be molecules in states $|\gamma\rangle$ composed of sublevels with $m=\pm2$, as shown by (\ref{MuStart}--\ref{MuEnd}). 

Following Eq. (\ref{Qexpt}), the quantity we measure is again
\begin{align}
\label{Qmix}
Q_\text{mix} = - \frac{1}{\tau P} \int \ln \left[ \frac{S_{2,m}(\delta)}{S_{2,m}(\infty)} \right] d\delta = \frac{A}{\tau P}.
\end{align}
Using (\ref{DefineR}--\ref{Qmix}), the correction factor is then
\begin{align}
\label{Rexplicit}
R(m_1, m_2) = - \, \frac{ \tau \int \Gamma_{m_1,m'} (\delta) \, d\delta}{ \int \ln \left[ {S_{2,m_2}(\delta)}/{S_{2,m_2}(\infty)} \right] d\delta }.
\end{align}
For the mixed quantization considered here, the signs of the arguments do not matter:  $R(m_1,m_2) = R(|m_1|,|m_2|)$ because $f_\gamma(-m) = f_\gamma(m)$.
Note that in the absence of mixed quantization we would be free to choose $f_\gamma(m) = \delta_{\gamma,m}$, which recovers
$R(m,m) = 1$.

We measured the data for transitions involving $(J=2,m_1=\pm1)$ using $m_2 = 1$ or $-1$.
For this case, the ratio
\begin{align}
\frac{S_{2,1}(\delta)}{S_{2,1}(\infty)}\Biggr\rvert_{m_1=\pm1} = \text{exp}\left[- \frac{1}{2} \Gamma_{\pm1,m'}(\delta) \tau \right],
\end{align}
using Eq. (\ref{fmu}) with Eqs. (\ref{MuStart}--\ref{MuEnd}) and (\ref{Smixed}--\ref{n2m}).
The correction factor (\ref{Rexplicit}) is then $R(1,1) = 2$, which we denoted previously by $R(m_1) = R(\pm1)$.

For transitions involving $(J=2,m_1=0)$ we used $m_2 = 0$.
In this case the ratio is bi-exponential, but we can approximate it as
\begin{align}
&\frac{S_{2,0}(\delta)}{S_{2,0}(\infty)}\Biggr\rvert_{m_1=0}
	= \frac{1}{4} \text{exp}\left[- \frac{1}{4} \Gamma_{0,m'}(\delta) \tau \right] \nonumber \\
	&+ \frac{3}{4} \text{exp}\left[- \frac{3}{4} \Gamma_{0,m'}(\delta) \tau \right]
	\approx \text{exp}\left[- \frac{3}{4} \Gamma_{0,m'}(\delta) \tau \right].
\end{align}
This correction factor (\ref{Rexplicit}) is then $R(0,0) \approx 4/3$, which we denoted previously by $R(0)$.

Finally, for transitions involving $(J=2, m_1 = \pm 2)$ we used $m_2 = 0$.
Again, we can approximate the ratio
\begin{align}
&\frac{S_{2,0}(\delta)}{S_{2,0}(\infty)}\Biggr\rvert_{|m_1|=2}
	= \frac{1}{4} \text{exp}\left[- \frac{3}{8} \Gamma_{\pm2,m'}(\delta) \tau \right] \nonumber \\
	&+ \frac{3}{4} \text{exp}\left[- \frac{1}{8} \Gamma_{\pm2,m'}(\delta) \tau \right]
	\approx \text{exp}\left[- \frac{1}{8} \Gamma_{\pm2,m'}(\delta) \tau \right].
\end{align}
This correction factor (\ref{Rexplicit}) is then $R(2,0) \approx 8$, which we denoted previously by $R(\pm2)$.

\subsection{Calculation of transition strengths}

The theoretical values for transition strengths are obtained by calculating the square of the transition dipole matrix element between a ground state $|\gamma\rangle$ and an excited state $|\mu(B)\rangle$ for an electric field linearly polarized along the magnetic field, $|\langle \gamma | d_Z | \mu(B)\rangle |^2$.  Here, $d_Z$ is the space-fixed transition dipole moment, which is expressed in terms of the molecule-fixed $d_i$ components by means of Wigner rotational matrices.  We assume that the starting level $|\gamma\rangle$ belongs to the electronic ground state X$^1\Sigma_g^+$, has well defined $J$, $m$, and $\Omega$ quantum numbers ($\Omega$ is the projection of the total electronic angular momentum along the internuclear axis), and do not include any corrections for the coupling of $m$ sublevels from mixed quantization by the optical lattice. The target level $|\mu(B)\rangle$ belongs to the $0_u^+/1_u$ potentials near the ${^1S_0} + {^3P_1}$ asymptote, and is a superposition of states with different $J'$ and $|\Omega'|$, as in Eq.~(1), such that only $m$ remains a good quantum number in the presence of the magnetic field.  The correct form of $|\mu(B)\rangle$ as a function of the field strength $B$ is obtained by solving the coupled-channel equations with the Zeeman Hamiltonian (which couples states with $\Delta J' = 0, \pm1$) and nonadiabatic Coriolis coupling (which couples $1_u$ and $0_u^+$ electronic states of the same $J'$) included.  The calculations include couplings between states from $J'=1$ up to $J'=6$, which results in nine coupled channels in total (each even $J'$ contributes a single $1_u$ channel, and each odd $J'$ contributes two ($1_u$ and $0_u^+$) channels).  We assumed that the molecule-fixed electronic transition dipole moment between the X$^1\Sigma_g^+$ and $0_u^+/1_u$ states is independent of the internuclear separation $R$ and equal to its asymptotic atomic limit,
\begin{align}
\langle \mathrm{X}{^1\Sigma}_\text{g}^+ | d_i | 1_\text{u}/0_\text{u}^+ \rangle \approx \sqrt{2} \langle ^1S_0| d_i | ^3P_1 \rangle,
\end{align}
which is a good approximation for weakly bound molecules \cite{mcguyer:1g}.
Just as with the experimental results, the reported values are normalized to the average zero-field strength of the chosen allowed transition in Fig.~2(c).


\end{document}